\documentclass[onecolumn]{article}

\usepackage{graphicx}
\usepackage{ amssymb}
\usepackage[ansinew]{inputenc}
\usepackage{authblk}
\RequirePackage{subfig}

\begin{document}

\title{Experimental evidence of planar channeling in a periodically bent crystal}

\author[1]{E. Bagli} 
\author[1]{L. Bandiera} 
\author[1]{V. Bellucci} 
\author[2]{A. Berra} 
\author[1]{R. Camattari} 
\author[3]{D. De Salvador} 
\author[1]{G. Germogli} 
\author[1]{V. Guidi} 
\author[4]{L. Lanzoni} 
\author[2]{D. Lietti} 
\author[1]{A. Mazzolari} 
\author[2]{M. Prest} 
\author[5]{V. V. Tikhomirov} 
\author[6]{E. Vallazza}

\affil[1]{INFN Sezione di Ferrara, Dipartimento di Fisica e Scienze della Terra, Universit{\`a} di Ferrara Via Saragat 1, 44122 Ferrara, Italy}
\affil[2]{Universit{\`a} dell'Insubria, via Valleggio 11, 22100 Como, Italy $\&$ INFN Sezione di Milano Bicocca, Piazza della Scienza 3, 20126 Milano, Italy }
\affil[3]{INFN Laboratori Nazionali di Legnaro, Viale dell'Universit{\`a} 2, 35020 Legnaro, Italy $\&$ Dipartimento di Fisica, Universit{\`a} di Padova, Via Marzolo 8, 35131 Padova, Italy}
\affil[4]{Department of Engineering, University of San Marino, Via Salita alla Rocca 44, 47890, Republic of San Marino}
\affil[5]{Research Institute for Nuclear Problems, Belarusian State University, Bobruiskaya street, 11, Minsk 220030, Belarus}
\affil[6]{INFN Sezione di Trieste, Via Valerio 2, 34127 Trieste, Italy }

\maketitle

\begin{abstract}
The usage of a Crystalline Undulator (CU) has been identified as a promising solution for generating powerful and monochromatic $\gamma$-rays. A CU was fabricated at SSL through the grooving method, i.e., by the manufacturing of a series of periodical grooves on the major surfaces of a crystal. The CU was extensively characterized both morphologically via optical interferometry at SSL and structurally via X-ray diffraction at ESRF. Then, it was finally tested for channeling with a 400 GeV/c proton beam at CERN. The experimental results were compared to Monte Carlo simulations. Evidence of planar channeling in the CU was firmly observed. Finally, the emission spectrum of the positron beam interacting with the CU was simulated for possible usage in currently existing facilities.
\end{abstract}

\section{Introduction}

High-intensity and monochromatic X-ray sources are important tools for research in fundamental to applied science.  Nowadays, intense and monochromatic soft X-ray beams are produced by means of Free-Electron Lasers (FELs). FELs are based on magnetic undulators \cite{Ginzburg,Motz}, which force electrons to an oscillatory motion, resulting in electromagnetic radiation generation. With currently available magnetic undulators, the minimum achievable oscillation period $\lambda_u$  is of the order of the centimeters, thus limiting the generation of X-ray to a few tens of keV at the highest synchrotron electron energies \cite{1325024}.

The availability of harder X-rays or even $\gamma$-ray source will pave the way to the development of innovative applications. For instance, a $\gamma$-ray beam can induce nuclear reactions through photo-transmutation \cite{nuclearph}, i.e., it can be employed for changing the atomic number of nuclei. This technique can be employed for eliminating nuclear waste by trasmuting it into short-lived nuclei for medicine as $^{126}$Sn$(\gamma,$n$)^{125}$Sn \cite{Irani2012466} or the $^{100}$Mo$(\gamma,$n$)^{99}$Mo reaction followed by a $\beta$-decay used to produce the $^{99m}$Tc isotope \cite{national2009Medical}. Another possible application lies in the field of photo-induced nuclear fission to induce the fragmentation of heavy nuclei \cite{PhysRev.56.426,PhysRev.59.57}. Supply of energy from the gamma-quanta causes the excitation of nuclei. Provided that the photon energy matches the transition energy between nuclear states, the reaction acquires a resonant character. As a possible result, heavy nuclei are splitted into two or more fragments. As an example of application, this process can be used in the production of medium mass neutron-rich nuclei.

In order to produce photon with the suitable energies for the applications, undulators with a shorter ${\lambda}_u$ than currently available FELs are needed. First efforts in this direction date back to the mid 80s \cite{PhysRevSTAB.15.070703}. A design for mm-scale pulsed electromagnetic undulator was proposed in Ref. \cite{Granatstein}, a mm-scale undulating field produced using periodic grooves ground into samarium cobalt blocks was proposed and obtained \cite{Ramian,Paulson}, and hybrid-bias-permanent-magnet undulators with period lengths in the range of ${\lambda}_u$ = $700 - 800$ $\mu$m were fabricated \cite{Tatchyn1,Tatchyn2,Tatchyn3}.By exploiting the recent progress in the fabrication of MicroElectroMechanical Systems (MEMS), the realization of undulators with ${\lambda}_u$ in the 10 - 1000 $\mu$m range has been proposed \cite{PhysRevSTAB.15.070703}. Drawbacks of the MEMS undulators are the demand for an expensive cooling system and the strict requirements on beam quality.
	
A promising solution for reaching higher photon energies is the usage of a Crystalline Undulator (CU) \cite{Baryshevsky198061,Baryshevsky2,Baryshevsky201330}. In a CU, the electrons (positrons) are forced to an oscillatory motion by the strong electrostatic field generated by the aligned atoms of a crystal planes or axes. For instance, the Si (110) planes act on charged particles with an electric field of 6 GV/cm, which is equivalent to a 2 kT magnetic field. Therefore, the trajectories of the particles entering a crystal at small angle with respect to the major crystallographic planes or axes, are confined \cite{Dansk.Fys.34.14}. The quasi-oscillatory motion of the channeled particles is accompanied by channeling radiation.

A crystal can tolerate a significant amount of stress and torsion while remaining into the elastic regime. Therefore, a complex geometry, such as the periodically bent crystal needed to drive the particles inside manipulated channels, can be achieved. By exploiting currently available techniques, crystals with an undulated geometry with a period ranging from 1 $\mu$m to 1 cm can be fabricated.

\begin{figure}
\begin{center}
\includegraphics[width=1\columnwidth]{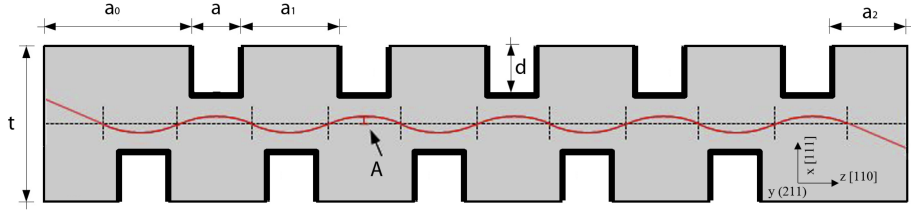}
\caption{Schematic representation of the crystalline undulator obtained through the grooving method. The red curved line represents the undulated planes in the center of the crystal. $\textit{A}$ is the amplitude of the undulated planes, $\textit{a}$ the groove width and $\textit{d}$ the groove depth.}\label{fig:schema}
\end{center}
\end{figure}

\section{Fabrication}
\begin{figure}
\begin{center}
\subfloat[3D reconstruction of the sample surface obtained through interferometric profilometry]{\includegraphics[width=0.8\columnwidth]{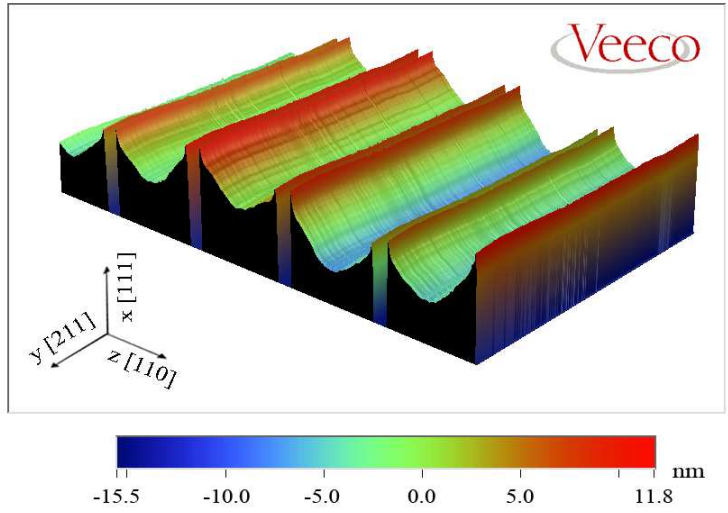}}
\newline
\subfloat[Angular analysis of (220) planes through X-ray diffraction.]{\includegraphics[width=0.8\columnwidth]{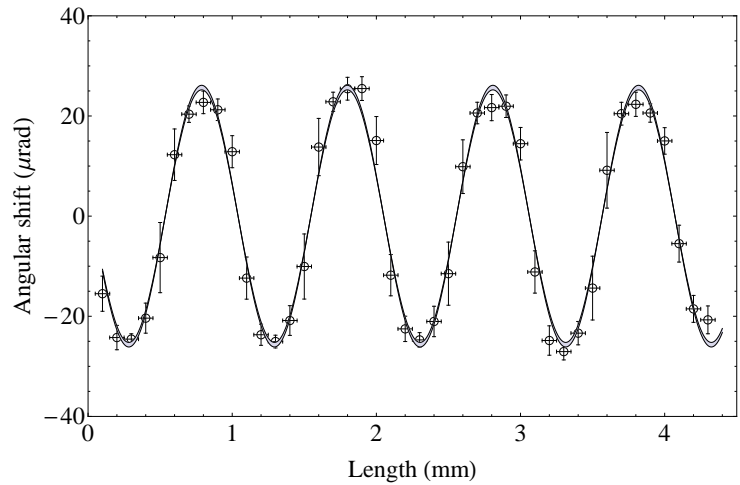}}
\caption{Morphological and structural measurement of the sample.}\label{fig:onduinterESRF}
\end{center}
\end{figure}
In order to manufacture a working CU, a crystal with very low density of defects should be used, together with high accuracy in bending the crystallographic planes. Indeed, the capture efficiency of channeling is strongly reduced by the presence of crystalline defects \cite{PhysRevLett.110.175502}, and the coherence of the emitted X- and $\gamma$-rays relies on the accuracy in the periodically bending of the crystallographic planes. Various methods have been proposed for the realization of a periodically bent crystal, such as acoustic-wave transmission \cite{Baryshevsky198061}, periodically-graded $Si_{1-x}Ge_x$ structures \cite{Breese1997540}, periodic-surface deformations obtainable via superficial grooves \cite{Afonin2005122,PhysRevSTAB.7.023501}, laser ablation \cite{Balling20092952}, or film deposition \cite{Guidi200540,Guidi2007apl}.

The CU radiation can be distinguished from the channeling radiation only if ${\lambda}_u$ is not equal to the average channeling oscillation period ${\lambda}_c$. Traditional CUs work under the condition ${\lambda}_u<{\lambda}_c$, though an alternative scheme with ${\lambda}_u>{\lambda}_c$ was recently proposed \cite{PhysRevLett.110.115503}.

Several experiments aimed to demonstrating the feasibility of a working CU for the traditional configuration \cite{PhysRevLett.90.034801,Afonin2005122,biryukovundu,Baranov200632,Baranov2005,Backe201337,1742-6596-438-1-012017} were performed, and some theoretical works were accomplished to foresee the energy generated by CUs \cite{korol2004,PhysRevLett.98.164801}. Recently, a first evidence of the radiation generated by a sub-GeV electrons interacting with a CU having ${\lambda}_u<{\lambda}_c$ was observed in the 7-15 MeV range, using a 600-855 MeV electron beam \cite{PhysRevLett.112.254801}.

The grooving method \cite{PhysRevLett.90.034801} may be a simple and reproducible solution to fabricate a CU that emits radiation in the MeV range working within the traditional ${\lambda}_u<{\lambda}_c$ scheme.

In this paper, we present the first experimental evidence of planar channeling in a CU fabricated via superficial grooves.

The superficial grooving method consists of making a series of grooves on the major surfaces of a crystal. In the framework of the LAUE project \cite{SPIE2013Dante}, such method has been deeply investigated and mastered to produce bent crystals tailored for the realization of optics for hard-X- and $\gamma$-rays, i.e. Laue lenses \cite{CamaQM,150Ge}. It was shown that a series of grooves may cause a permanent and reproducible deformation of the whole sample \cite{indentazioni1,indentazioni2}. Indeed, the plasticization that occurs in the thin superficial layer transfers co-active forces to the crystal bulk, producing an elastic strain field within the crystal itself. Making an alternate pattern of parallel grooves on both surfaces of a crystal, the realization of a millimetric or even a sub-millimetric undulator could be envisaged.

The CU described in the present paper was realized by making a series of grooves on a 0.2$\times$45.0$\times$5.0 mm$^3$ silicon crystal. In particular, an alternate and periodic pattern of  $150$ $\mu$m-wide parallel grooves was plotted on the largest surfaces of the strip, with the distance between consecutive grooves being 1 mm. Grooves were carved by means of a diamond blade through the usage of a DISCO DAD 3220 dicing machine. The groove depth was ($54\pm2$) $\mu$m. A schematic representation of the sample is illustrated in Fig.\ref{fig:schema}.

The grooving parameters used for the CU fabrication were employed as input parameters to compute the expected amplitude of the undulated planes. In the framework of the classical theory of elasticity, the Stoney formula can be extended to evaluate the curvature just beneath each groove and, in turn, the undulator amplitude  \cite{indentazioni2}. The undulator amplitude resulted to be $(4.75\pm0.50)$ nm.

\section{Interferometric and diffractometric characterization}

A morphological analysis of the sample surface was attained at the Sensor and Semiconductor Laboratory (SSL) of Ferrara (Italy) through a VEECO NT-1100 white-light interferometer, which allowed the reconstruction of the three-dimensional surfaces with lateral resolution of about 1 $\mu$m and vertical resolution of 1 nm. The profilometric pattern of the sample surface (shown in Fig. \ref{fig:onduinterESRF}b), allows the estimation of the average undulator amplitude, which resulted to be ($4.5\pm1.0$) nm.

The CU was also tested through X-ray diffraction at the ID15A beamline of the European Synchrotron Radiation Facility (ESRF, Grenoble, France). A highly monochromatic beam was set at 150 keV, the beam size was 50$\times$50$ \mu$m$^2$. The sample was characterized by recording the position of the Bragg peak for the (220) lattice plane for 50 consecutive points along the [110] direction. In order to subtract a possible shift due to instrumental tolerances, a flat reference crystal with identical lattice orientation was set behind the undulator, and the (220) Bragg peak of the reference crystal was recorded together with the Bragg peak of the undulator. The shift of the angular position of the CU Bragg peak with respect to the reference as a function of the impact position of the X-ray beam is shown in Fig. \ref{fig:onduinterESRF}. The measured average amplitude of undulation was ($4.1\pm0.5$) nm with an average period of ($1.01\pm0.03$) mm.

\begin{figure}
\begin{center}
\includegraphics[width=0.8\columnwidth]{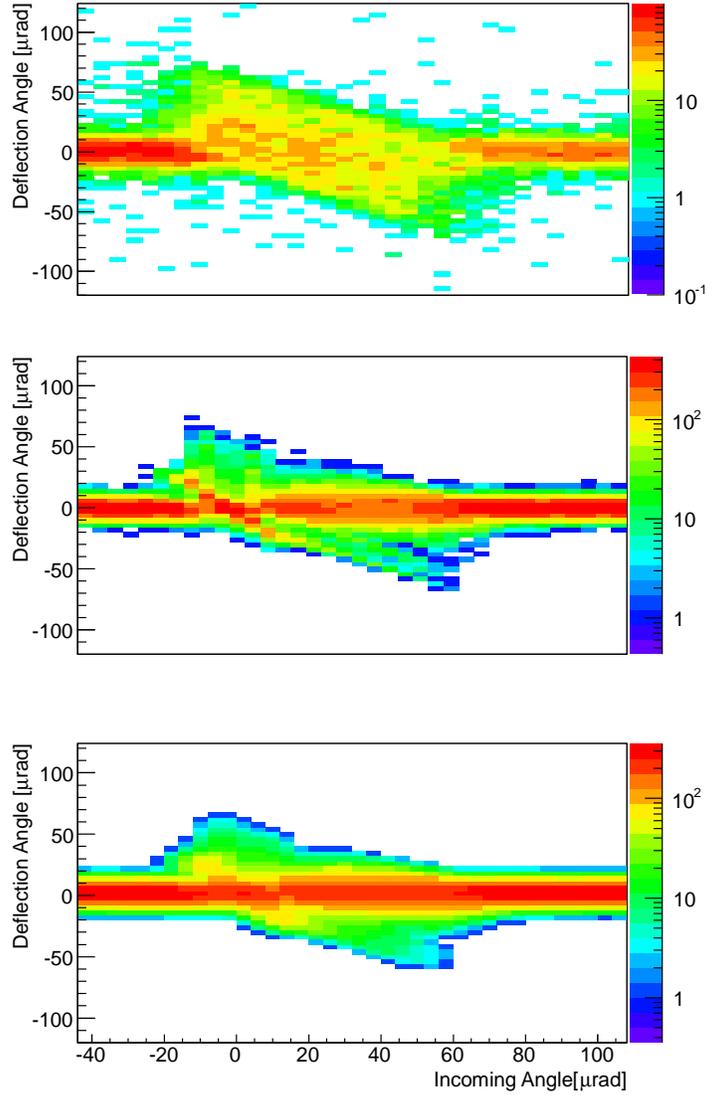}
\caption{(a) Measured distribution of outgoing particles after the interaction with the undulator as a function of the incoming angle and the deflection angle with respect to the crystal plane. (b) The Monte Carlo simulation of the same distribution. (c) The Monte Carlo simulation of the same distribution convoluted with the experimental resolution.}\label{fig:h8exp}
\end{center}
\end{figure}
\section{On-beam characterization}

The CU was exposed to a 400 GeV/c proton beam in the H8 beam line at CERN-SPS. It was mounted on a two-axis rotational stage with 2 $\mu$rad resolution. The beam was tracked before and after the interaction with the CU by using a telescope system of Si strip detectors \cite{Celano199649,PhysRevLett.101.234801}. The angular resolution of the tracker is $\sim3.5$ $\mu$rad \cite{HasanPhDThesis}. The beam size was ($1.36\pm0.02$) mm $\times$ ($0.73\pm0.01$) mm and the angular divergence ($10.15\pm0.04$) $\mu$rad $\times$ ($8.00\pm0.03$) $\mu$rad, as measured with the telescope system.

Remaining in the \emph{xz} plane, the \emph{z} axis of the CU was purposely misaligned with respect to the beam direction to avoid the undesired axial-channeling effect. The beam enters the crystal through a (110) oriented surface with an area of 0.2$\times$45.0 mm$^2$. The (111) planes were exploited as channeling planes.

The distribution of the deflection angle of the particles as a function of the horizontal incoming angle was measured by rotating the goniometer around the channeling position (see Fig. \ref{fig:h8exp}). Particles channeled along the whole crystal length acquire a null deflection angle. On the contrary, if a channeled particle undergoes dechanneling, the trajectory remains parallel to the tangent of the local plane curvature where dechanneling has occurred. As a consequence, particles may acquire either a positive or negative deflection angle. By comparing the experimental pattern with the Si (110) angular scan from the literature (see Figs. 4, 6, and 7 of \cite{PhysRevSTAB.11.063501}), a noticeable difference holds. In fact, the particles channeled in a uniformly bent crystal are deflected only to one among the two possible (positive or negative) directions.

\begin{figure*}
\begin{center}
\includegraphics[width=0.8\textwidth]{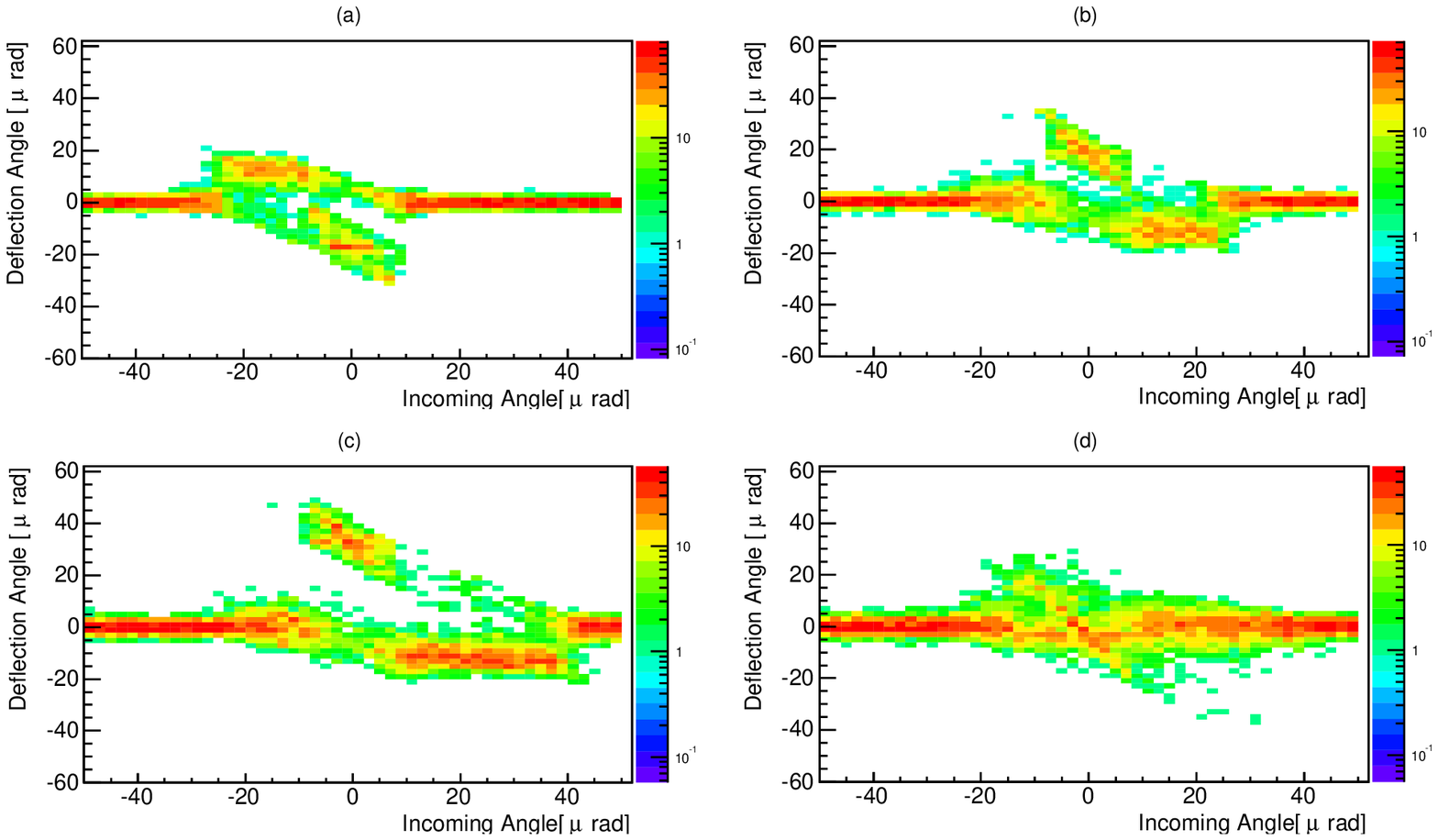}
\caption{Simulated distribution of particles during the interaction with the undulator as a function of the incoming angle and the deflection angle with respect to the crystal plane at four different penetration depth, i.e., 250 $\mu$m (a), 500 $\mu$m (b), 750 $\mu$m (c) and 1000 $\mu$m (d).}\label{fig:h8MC}
\end{center}
\end{figure*}

\section{Monte Carlo simulations}

In order to get an insight in the experimental result, a Monte Carlo simulation was performed through the DYNECHARM++ toolkit \cite{Bagli2013124,DYNECHARM_PHI}. The code solved the equation of motion in the non-inertial reference frame orthogonal to the crystal plane via numerical integration under the continuum potential approximation \cite{Dansk.Fys.34.14}. The electrical characteristics of the crystals were evaluated through the ECHARM software \cite{PhysRevE.81.026708}, which made use of diffraction data for the computation of the atomic form factors. The CU geometry was approximated through a sinusoidal function, with the amplitude being measured at ESRF. The oscillation period is taken equal to the groove distance, i.e. 1 mm. The influence of such geometry on particle trajectory was evaluated through the application of a position-dependent centrifugal force. Intensity of scattering on nuclei and electrons was averaged over the distribution density in the channel in order to reliably estimate the dechanneling probability, i.e., the probability for a channeled particle to leave the channeling condition.

Figure \ref{fig:h8MC} shows the distribution of the horizontal deflection angle as a function of the horizontal incoming angle at four different penetration depths. As shown in Fig. \ref{fig:h8MC}a, at the penetration depth of 250 $\mu$m, the clear typical markers of a bent crystal are present, namely the channeling spot and the volume reflected particles. In fact, channeled particles can be distinguished due to the $+20$ $\mu$rad deflection kick they received. On the other side, the particles that impinge the target with a direction tangent to the crystalline curved planes are deflected to the opposite direction, i.e., they undergo volume reflection (VR). In Fig. \ref{fig:h8MC}b, at the penetration depth of 500 $\mu$m, the channeled particles are deflected to the opposite angle and acquire a $-20$ $\mu$rad kick. Particles that underwent the VR effect are reflected by the opposite curvature. Thus, a null average deflection is observed. Particles pertaining to the region at the left of the channeling spot are also reflected. Then, Fig. \ref{fig:h8MC}c shows a shift of the channeling spot of $-35$ $\mu$rad at the 750 $\mu$m depth. Lastly, in Fig. \ref{fig:h8MC}d, at the 1000 $\mu$m depth, the general picture evolves towards a double fold distribution with a net overall null average deflection angle. The simulated pattern of the beam interacting with the whole CU is shown in Fig. \ref{fig:h8exp}b. Figure \ref{fig:h8exp}c shows the simulation convoluted with the detector angular resolution. The simulation is in good agreement with the experimental results.

With the same Monte Carlo code, the fraction of planar-channeled particles along the crystal length, i.e., the channeling efficiency, can be calculated. For a 400 GeV/c proton beam entering the CU parallel to the crystalline planes the efficiency  is $(48 \pm 1) \%$. Such quantity can also be estimated with an analytical model usually adopted for uniformly-bent crystals \cite{doi.org/10.1140/epjc/s10052-014-2740-7}. By setting the bending radius equal to the $6.2$ m minimum curvature of the planes in the CU, the efficiency results to be $\sim$ 60$\%$ for a zero-divergence beam.

The chosen CU sample parameters fulfill the condition for an optimal undulator in the case of 15 GeV positrons. At these energies, with this CU, the possibility to use an electron beam instead of a positron beam has to be excluded because the dechanneling length is much smaller than the CU length \cite{Scandale201370}. Thereby, only positrons can be used for such a test. In the next future, a positron beam in the energy range of 10-20 GeV will be available at FACET-SLAC. As done for the 400 GeV/c proton beam, the expected efficiency for 15 GeV/c positrons is higher than $40 \%$, so that the CU described in this paper could be applied straight in that facility.

A prediction regarding the electromagnetic radiation emitted by a 15 GeV positron beam passing through the CU sample was done through a specific Monte Carlo code \cite{Baryshevsky201330}. Since the undulator radiation is expected in the $1$ MeV region, the Ter-Mikaelian density effect and the transition radiation were also taken into account. The beam divergence was set to zero. The radiation emission probability was averaged over a large number of particle trajectories and photon emission angles.

\begin{figure}
\begin{center}
\includegraphics[width=1\columnwidth]{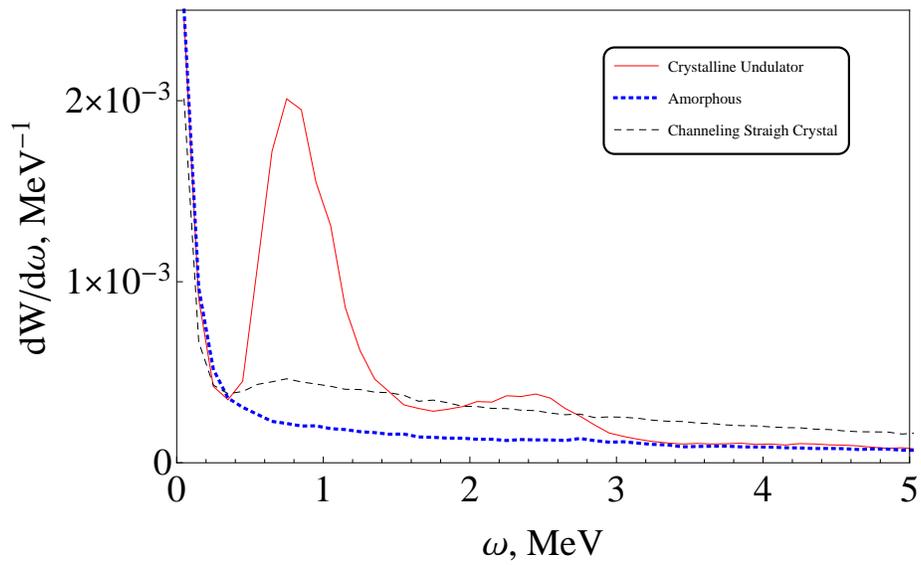}
\caption{Simulated radiation emission probability for 15 GeV positrons interacting with the CU sample presented in the paper (solid line) and with a straight Si crystal having the same length along the beam under channeling condition (dashed line). The simulated spectrum emitted by 15 GeV positrons interacting with an amorphous Si target of the same length (dotted curve) is also reported.}\label{fig:Rad}
\end{center}
\end{figure}

The distribution of the radiation emission probability generated by 15 GeV positrons interacting with the CU sample is shown in Fig. \ref{fig:Rad}. The emission spectrum has a sharp peak at $\sim$ 0.8 MeV. Because the channeling radiation (CR) is expected in the 100 MeV region \cite{BaierKatkov}, the main CR and CU peaks present no overlap. In the energy range from 0.5 to 1.5 MeV the peak of the CU radiation is $\sim$4 times higher than for CR.

\section{Conclusions}

A crystalline undulator to be worked out under the condition ${\lambda}_u<{\lambda}_c$ was manufactured through the grooving method at SSL (Ferrara, It). The sample was extensively characterized both morphologically and structurally. The surfaces were reconstructed through an interferometric measurement at SSL. The crystalline structure was analyzed through monochromatic hard X-rays diffraction at ESRF (Grenoble, Fr). The crystal was tested at the H8-SPS line of CERN. The expected planar channeling of 400 GeV/c protons between the cristalline planes bent with the grooving method was observed. Monte Carlo simulations were worked out to evaluate the channeling efficiency for the 400 GeV proton beam and to predict the radiation spectrum emitted by 15 GeV positrons beam entering the crystalline undulator. The manufactured crystalline undulator is now ready to be tested with high intensity positrons beam at 10-20 GeV energy.

\section{Acknowledgements}
We are thankful to the CERN-SPS and ESRF coordination board for the possibility to use the H8-SPS external line at CERN and the ID15A line at ESRF. We acknowledge the partial support of INFN under the ICE-RAD experiment and the European CUTE Project. We acknowledge Gerald Klug and Eugen Eurich from Disco Europe (Munich, Germany) for their support in crystal manufacturing, Persiani Andrea and Manfredi Claudio of Perman (Loiano, Italy) for their support with crystal holders manufacturing.

\bibliographystyle{unsrt}
\bibliography{biblio}

\end{document}